\documentclass[twocolumn,10pt,dvipsnames]{autart}    
\usepackage{amsmath,amssymb,xcolor}
\usepackage{bm}

\newtheorem{hyp}{\bf Assumption}

\newtheorem{lemma}{\bf Lemma}
\newtheorem{corollary}{\bf Corollary}
\newtheorem{definition}{\bf Definition}

\usepackage{tikz}
\usepackage{here}
\usetikzlibrary{arrows,shapes,calc}
\usepackage{pgfplots}
\usetikzlibrary{positioning} 

\usepackage{times}
\usepackage{enumitem}

\tikzstyle{solid}=                   [dash pattern=]
\tikzstyle{dotted}=                  [dash pattern=on \pgflinewidth off 2pt]
\tikzstyle{densely dotted}=          [dash pattern=on \pgflinewidth off 1pt]
\tikzstyle{loosely dotted}=          [dash pattern=on \pgflinewidth off 4pt]
\tikzstyle{dashed}=                  [dash pattern=on 3pt off 3pt]
\tikzstyle{densely dashed}=          [dash pattern=on 3pt off 2pt]
\tikzstyle{loosely dashed}=          [dash pattern=on 3pt off 6pt]
\tikzstyle{dashdotted}=              [dash pattern=on 3pt off 2pt on \the\pgflinewidth off 2pt]
\tikzstyle{densely dashdotted}=      [dash pattern=on 3pt off 1pt on \the\pgflinewidth off 1pt]
\tikzstyle{loosely dashdotted}=      [dash pattern=on 3pt off 4pt on \the\pgflinewidth off 4pt]

\usepackage{graphicx}      
\newcommand{\rev}[1]{{\color{black} #1}}

\begin{document}

\begin{frontmatter}

\title{A new formulation of Economic Model Predictive Control without terminal constraint} 


\author[Mazen]{Mazen Alamir}\ead{mazen.alamir@grenoble-inp.fr},  
\author[Gabriele]{Gabriele Pannocchia}\ead{gabriele.pannocchia@unipi.it},  

\address[Mazen]{Univ. Grenoble Alpes, CNRS, Grenoble INP, GIPSA-lab, 38000 Grenoble, France.}
\address[Gabriele]{Dept. of Civil and Industrial Engineering, University of Pisa, 56126 Pisa, Italy}  


\begin{abstract}                          
In this paper, it is shown that a simple formulation of Economic Model Predictive Control can be used which possesses two features that are generally viewed as mutually exclusive, namely, a rather short prediction horizon (reachability-compatible) on one side, and the absence of final constraint on the other side. Practical stability at an arbitrarily small neighborhood of the optimal unknown steady-state pair is shown when some design parameters increase. It is also shown that when the system is originated from the time discretization of a continuous-time dynamics, the size of the terminal region can be reduced by decreasing the sampling period for the same design parameter setting. A commonly used example is given to illustrate the results. 
\end{abstract}

\end{frontmatter}

\section{Introduction} \label{secintro}
Model Predictive Control (MPC) refers to a wide class of optimization-based control methods in which a dynamic process model and numerical optimization are exploited to evaluate, repeatedly online, an optimal sequence of state (output) and input trajectories; in closed-loop operation the first portion of the control sequence is actually implemented into the plant, and the overall planning is repeated at the next sample time~\cite{rawlings:mayne:diehl:2017}. 
Economic Model Predictive Control (EMPC) formulations~\cite{Rawlings09b,Amrit:11,Diehl:11a} are MPC formulations in which the cost function is not expressed as a distance to some beforehand known targeted steady pairs. In a typical non economic settings, such pairs are computed in a higher stage of a two stage formulation in which the lower stage is a standard regulation-based MPC -with preassigned targeted steady state- while the higher stage performs economic static optimization in order to deliver an optimal steady pair to the lower stage~\cite{qin:badgwell:2003}.

The advantage of EMPC stems from the fact that there exist (non equilibrium) input/state pairs for which the economic cost is lower than that corresponding to the equilibrium target, so that transient operation away from the optimal equilibrium can be more remunerative than reaching the equilibrium target ``quickly'' as in conventional two-layer architectures~\cite{Rawlings09b}.
Moreover, the computation of steady-state targets may be also time consuming, thus possibly avoidable.
In some cases, it is even possible that a non necessarily steady behavior induces a higher performance on average~\cite{Angeli:12a}.
In both cases however, it is necessary to be able to \rev{limit and in the long term to suppress the movements of actuators in order to achieve a quasi-optimal steady regime as a limit case.}
\rev{As a matter of fact, some approaches include a modification of the economic cost function by adding a tracking term (based on the presumably known optimal equilibrium) \cite{amrit2013,maree2016}; this can also be interpreted within a multi-objective framework \cite{zavala2015}.}


The discussion above suggests that penalizing the state increment should be effective in deriving a tunable EMPC that addresses the above concerns, \rev{as this penalty induces convergence towards an equilibrium.}
Strangely enough this simple and intuitive idea never showed up in any of the yet developed provable EMPC schemes to the best of the authors' knowledge. Instead, the less intuitive concepts of (strict) passivity, stage cost rotation and so, dominated the scene. This paper aims to fill this gap by giving a simple provably stable formulation that does not need these technicalities.  

On the other hand, existing provably stable formulations are mainly of two kinds. 
In the first, a terminal constraint and/or terminal penalty, based on the distance between the terminal state and the equilibrium target, is added~\cite{Diehl:11a,Angeli:12a}, possibly with an additional tracking term added to the cost function ~\cite{Griffith17a}, which undermines one of the attractive features of EMPC mentioned above (the \rev{non-availability} of the steady-optimal pair).
In the second, the stability argument relies on the prediction horizon being sufficiently \rev{high} \cite{Gruene13a}  which can be computationally expensive. 
Recent formulations \cite{kit:zanon18a,faulwasser:pannocchia:2019} are somehow in the middle field of these approaches, as they avoid the use of terminal constraints by adding gradient correcting end penalties, still requiring either a sufficiently long prediction horizon or solving the steady-state optimal problem.


The framework proposed in the present contribution gathers the nice properties of both, namely, the possibility to use a moderate prediction horizon (the one linked to the reachability assumption) while being free of any terminal constraint or any knowledge of the optimal equilibrium state. 
This paper is organized as follows: first of all, the problem is stated and notation is introduced in Section \ref{probstatnot}. The working assumptions that are needed to derive the main result are given in Section \ref{sec-WA}. Section \ref{seccla} gives the statement and the proofs of the main results before an illustrative example is proposed in Section \ref{secExamples}.

\section{Problem statement and notation} \label{probstatnot}
Consider general nonlinear systems governed by the following discrete-time dynamics:
\begin{equation}
x^+=f(x,u) \qquad (x,u)\in \mathbb{R}^{n}\times \mathbb{R}^{m} \label{dynamics}
\end{equation}
where $x$ and $u\in \mathbb{U}\subset \mathbb{R}^{m}$ stand for the state and the control input vectors respectively. $\mathbb U$ is a compact set of admissible control values. 

When the dynamics (\ref{dynamics}) is obtained by time discretization of some continuous-time dynamics, the corresponding sampling period will be denoted by $\tau>0$. Otherwise, $\tau=1$ might be used every time $\tau$ is involved in the sequel. 

The dynamics is supposed to admit a set $\mathcal Z\subset \mathbb{R}^{n}\times \mathbb U$ of equilibrium pairs $(x,u)$, namely:
\begin{equation}
\mathcal Z:= \Bigl\{(x,u)\in \mathbb{R}^{n}\times \mathbb U\ \vert \ \Delta(x,u)=0\Bigr\} \label{defdecalZ}
\end{equation}
where $\Delta(x,u):=\dfrac{1}{\tau}\|f(x,u)-x\|$. Note that although $\Delta$ depends on $\tau$, the reference to $\tau$ is omitted for the sake of simplification. This dependence will be recalled when appropriate. 

Finally, it is assumed that there exists at least an optimal equilibrium pair $z_s=(x_s,u_s)\in \mathcal Z$ that minimizes a given cost function $\ell$ over the set of steady pairs $\mathcal Z$, namely:
\begin{equation}
\ell_s:=\ell(z_s)\le \ell(z)\quad \forall z\in \mathcal Z \label{condavanr}
\end{equation}
More generally the set of such $z_s$ is denoted by $\mathcal Z_s\subset \mathcal Z$. 

In what follows the following notation is used:
\begin{itemize}
\item let \rev{$N\in \mathbb{Z}_{+}$} be some finite prediction horizon
\item boldfaced $\bm u$ denotes a sequence of $N+1$ control actions over a prediction horizon of length $N$, namely:
\begin{equation}
\bm u:=(u_0,u_{1},\dots, u_{N-1}, u_{N}) \in \mathbb U^{N+1}\label{defdeubf}
\end{equation}
\item For any control sequence $\bm u$ given by (\ref{defdeubf}), the following notation is used to denote the corresponding {\em warm start} sequence:
\begin{equation}
\bm u^+ = (u_1,u_{2},\dots, u_{N}, u_{N})\in \mathbb U^{N+1} \label{ubfdag}
\end{equation}
\item Given a control sequence $\bm u\in \mathbb U^{N+1}$ and an initial state $x$, $\bm x^{\bm u}(x):=\{x^{\bm u}_k(x)\}_{k=0}^N$ denotes the sequence of states on the system's trajectory starting at $x$ under the control sequence $\bm u$, namely\footnote{Note that the subscript $k$ in $x^{\bm u}_k(x)$ denotes time increment and not state vector-related component index.}:
\begin{equation}
x^{\bm u}_0(x)=x ,  \quad x^{\bm u}_{k+1}(x)=f(x^{\bm u}_k(x),u_k) \quad \rev{\forall k}
\end{equation}
\item Given any function $h$ defined on $\mathbb{R}^{n}\times \mathbb U$, the following short notation is used:
\begin{equation}
h^{\bm u}_k(x):=h(x^{\bm u}_k(x),u_k)
\end{equation}
This holds in particular for $\ell$ and $\Delta$ invoked earlier. Moreover, when there is no ambiguity regarding the initial state $x$, the argument $x$ is omitted leading to the notation $x^{\bm u}_k, h^{\bm u}_k$ instead of $x^{\bm u}_k(x), h^{\bm u}_k(x)$. \\
\item For any set $\mathbb V$ in some euclidian space, an $\epsilon$-neighborhood of $\mathbb V$, denoted by $\mathcal V_\epsilon(\mathbb V)$ is the set of all points $\xi$ such that \rev{$d(\xi,\mathbb V)\le \epsilon$, in which $d(\xi,\mathbb V):=\min_{\eta\in \mathbb V} \|\xi-\eta\|$ is the distance of $\xi$ from $\mathbb V$.}
\end{itemize}
The following definition is used to express the main convergence results:
\begin{definition}[Quasi-steady optimal trajectory] \label{defregimquasi}
A trajectory $(\bm x, \bm u)$ is said to be $\epsilon$-quasi steady optimal if and only if the following conditions hold for all $k$:
\begin{align}
&\vert \ell(x_k, u_k)-\ell_s\vert \le \epsilon \quad,\quad \Delta(x_k, u_k)\le \epsilon\label{quasi}
\end{align}
\end{definition}
Using the above notation, the following open-loop cost function is considered in the present paper for a given initial state $x$ and a candidate sequence of future actions $\bm u$ over a prediction horizon of length $N$:
\begin{equation}
J(\bm u,x):=\gamma\underbrace{\Bigl[\ell_N^{\bm u}+\alpha\Delta^{\bm u}_N\Bigr]}_{\Psi_N(\bm u,x)}+\underbrace{\sum_{k=0}^{N-1}\Bigl[\ell_k^{\bm u}+\alpha\Delta^{\bm u}_k\Bigr]}_{V(\bm u,x)} \label{LaCost}
\end{equation}
where the notation $h^{\bm u}_k:=h^{\bm u}_k(x)$ is used for $h\in \{\ell, \Delta\}$.
We note that:
\begin{equation}\label{eq:Deltak}
\Delta^{\bm u}_k := \Delta (x^{\bm u}_k, u_k) = \dfrac{1}{\tau}\|f(x^{\bm u}_k,u_k)-x^{\bm u}_k\| = \dfrac{1}{\tau}\| x^{\bm u}_{k+1}-x^{\bm u}_k\| 
\end{equation}
This cost function is used to define the open-loop optimal control problem given by:
\begin{equation}
P(x): \min_{\bm u\in \mathbb U^{N+1}}J(\bm u,x)\rightarrow (J^\star(x), \bm u^\star(x)) \label{defdePx}
\end{equation}
leading to the MPC state feedback given by:
\begin{equation}
\kappa_{\mbox{\rm \tiny MPC}}(x):= u^\star_0(x) \label{feedback}
\end{equation}
We are interested in so called Economic-MPC formulations, meaning that $\ell(x,u)$ is not  defined as a distance to some desired equilibrium pair $(x_s,u_s)\in \mathcal Z_s$ that minimizes $\ell$ as the latter is supposed to be unknown or its a priori on-line computation is to be avoided. Consequently, no reference to such pair is included in the cost function nor in any terminal constraint. \\ \ \\ 
This paper investigates the conditions under which the resulting closed-loop behavior of (\ref{dynamics}) under (\ref{feedback}) is asymptotically $\epsilon$-quasi steady optimal with an $\epsilon$ that can be made as small as desired by conveniently choosing the design parameters $(\alpha,\gamma)$.\\ \ \\ 
In what follows, the following short notation is used:
\begin{equation}
h^\star_k(x):= h^{\bm u^\star(x)}_k(x)\qquad h\in \{\ell, \Delta\} \label{notation*}
\end{equation}
\begin{rem}
Note that only control constraints are explicitly considered through the subset $\mathbb U$. State constraints are assumed to be softened through some exact penalty in the definition of the map $\ell$. This greatly simplifies the exposition of the main ideas and avoid additional technicalities regarding the recursive feasibility issue. Moreover, it is actually the commonly used approach in real-life problems. {\color{black} A brief discussion regarding the possible constraint violation issue and the impact of the choice of the penalty on the exact constraint penalty is proposed in Section \ref{secsoft}.}
\end{rem}
\section{Working assumptions}\label{sec-WA}
\begin{hyp}[Properties of $f$ and $\ell$]\label{ellbounded} 
The following conditions hold true.
\begin{enumerate}[label=(\roman*)]
\item  $f$ and $\ell$ are continuous.
\item $\forall \bar\ell>0$, the set $B_{\bar \ell}:=\{x\in \mathbb{R}^{n}$ s.t. $\exists u\in \mathbb U, \ell(x,u)\le \bar\ell\}$ is compact.
\item The minimal steady value invoked in (\ref{condavanr}) is $\ell_s=0$.
\end{enumerate}
\end{hyp}
Assumption \ref{ellbounded}-(ii) is typically enforced by the soft constraints-related penalty [such as $\max\{0,\underline x-x, x-\bar x\}\le 0$] that is included in the definition of $\ell$ to express constraints on the state evolution. As for the last assumption 1-(iii), it is a standard assumption that is commonly introduced without loss of generality in order to simplify the discussion. 
\begin{hyp}[$N$-reachability of steady optimal pair]\label{Ass_reach} 
There exists a set $\mathbb X_0\subset \mathbb{R}^{n}$ such that for any $x\in \mathbb X_0$, $\exists \bm u^\#(x)\in \mathbb U^{N+1}$ s.t. $(x^{\bm u^\#}_N(x),u^\#_N(x))\in \mathcal Z_s$.
\end{hyp}
This is a standard assumption that is used in the convergence proof of MPC schemes. Note that the knowledge of the control sequence $\bm u^\#(x)$ is not required. Only its existence is needed for the analysis of optimal solution properties. 
\begin{hyp}[Optimal stationarity condition]\label{hyppsi}
There exists a class-$\mathcal{K}$ function $\psi(\cdot)$ such that:
\begin{equation}
0=\ell_s\le \ell(z)+\psi(\Delta(z)) \label{psi}
\end{equation}
Moreover, $\psi$ is Lipschitz-continuous.
\end{hyp}
Note that this last assumption simply states that for any stationary pair $z = (x, u)$, since $\Delta(z)=0$, there holds $\ell_s\le \ell(z)$. This {\color{black} statement includes} a simple technical rewriting of condition (\ref{condavanr}) {\color{black} together with a Lipschitz continuity requirement}. 
It also means that $\ell(z)$ might be lower than $\ell_s$ provided that $z$ is not a stationary pair. Recall that since $\ell_s$ is supposed to be $0$ (Assumption \ref{ellbounded}), the inequality (\ref{psi}) becomes $0\le \ell(z)+\psi(\Delta(z))$.
\begin{hyp}[Local properties] \label{localAss}
There exists a continuous function $K_s$, vanishing at zero, such that, for any $z:=(x,u) \in \mathbb{X}_0\times \mathbb U$, the following implication holds:
\begin{equation}
\Bigl\{d(z,\mathcal Z_s)\le \epsilon\Bigr\}\ \Rightarrow\ V(\bm u^{\#}(x),x)\le K_s(\epsilon)
\end{equation}
where $V$ is the integral cost defined in (\ref{LaCost}).
\end{hyp}
This is a rather weak technical assumption which can be expressed in simple terms as follows: when the state is in the neighborhood of $\mathcal Z_s$, it can be steered inside $\mathcal Z_s$ with low cost which vanishes with the initial distance to $\mathcal Z_s$.  Note that the knowledge of $K_s(\cdot)$ is not required, only its existence is needed to prove the main result. \\ \ \\ 
The last assumption that is needed to derive the main result is the following:
\begin{hyp} \label{recfeas}
For each $(x, \mathbf{u}) \in \mathbb{X}_0 \times \mathbb{U}^{N+1}$, the scalar map defined by
\begin{equation}
\Psi_N (\bm u,x):= [\ell+\alpha\Delta](x^{\bm u}_N(x),u_N)
\end{equation}
satisfies the following implication for sufficiently small $\vert \eta\vert$:
\begin{equation}
	\Bigl\{\Psi_N (\bm u,x)=\eta\Bigr\} \ \Rightarrow \Bigl\{\exists \bm u^\dag \in \mathbb{U}^{N+1} \ \vert\ \Psi_N(\bm u^\dag,x)=0\Bigr\} \label{condcindAss}
\end{equation}
\end{hyp}
\begin{rem}\label{comonAssfin}
Assumption~\ref{recfeas} is obviously very difficult to check. Nevertheless, we believe that it is not so restrictive in practice. The following comments can shed some light on the relevance and the restriction it implies:
\begin{enumerate} 
\item  Note first of all that for differentiable setting and in the absence of saturation on the control, Assumption \ref{recfeas} can be expressed in terms of the implicit function theorem, namely, a slight modification of $\eta$ into $0$ can be {\em compensated} by a corresponding slight modification in the argument $\bm u$. For this to hold, it suffices that the rank of the sensitivity of $\Psi_N$ to $\bm u$ at $(x,\eta)$ be equal to $1$ which is generically true.\\
\item When constraints are involved, satisfaction of the condition is no more trivial. Still, assuming that the sequence $\bm u:=(\bm u^{(a)}, \bm u^{(na)})$ satisfying $\Psi_N(\bm u,x)=\eta$ can be split into saturated components $\bm u^{(a)}$ and non saturated components $\bm u^{(na)}$. In this context, Assumption \ref{recfeas} suggests that the sensitivity of $\Psi_N$ to $\bm u^{(na)}$ is of rank 1 so that one can always perturb $\bm u^{(na)}$, keeping unchanged $\bm u^{(a)}$ in order to compensate for the infinitesimal change on $\Psi_N$ induced by infinitesimal $\eta$.\\
\item Note that the conditions of Assumption \ref{recfeas} could have been required only on the optimal sequences $\bm u^\star(x)$ rather than on any sequence of control $\bm u$. This is because in the sequel, the implication in (\ref{condcindAss}) is only used for such optimal sequences. \\
\item Based on the above discussion, it comes out that Assumption \ref{recfeas} can be replaced by several other checkable Assumptions of low level. But this might induce unnecessary conditions such as differentiability while these conditions are only sufficient. It is preferred here to keep the high level condition (\ref{condcindAss}) that might hold even for non differentiable settings. The above discussion helps for better understanding the underlying requirements. 

\end{enumerate}
\end{rem}
\section{Closed-loop analysis}\label{seccla}
We start by establishing a result that builds a first bridge between the penalty $\ell+\alpha\Delta$ used in the cost function (\ref{LaCost}) and the property ($\ell=0$ and $\Delta=0$) of optimal steady pairs. 
\begin{lemma}\label{lemmmmm}
Given a compact set $\mathbb X\times \mathbb U$, let $L_\psi$ be the Lipschitz constant of $\psi (\Delta(x,u))$ over $z:=(x,u) \in \mathbb X\times \mathbb U$. 
For any $\alpha>L_\psi$ the following implication holds true: 
\begin{equation}
\ell(z)+\alpha \Delta(z)\le \epsilon \quad \Rightarrow\quad  \left\{ 
\begin{array}{l}
 \Delta(z)\le \kappa_1\epsilon\\
 \ell(z)\in [-\kappa_2\epsilon,\epsilon]
\end{array}
\right. \label{lachose}
\end{equation}
where $\kappa_1=1/(\alpha-L_\psi)$ and $\kappa_2=\alpha/(\alpha-L_\psi)$.
\end{lemma}
{\sc Proof}. Using the inequality (\ref{psi}) of Assumption \ref{hyppsi}, and recalling that $\ell_s=0$ is used without loss of generality, it follows that:
\begin{equation}
\ell(z)\ge -\psi(\Delta(z))
\end{equation}
combining this with the left hand side of (\ref{lachose}) leads to the following inequality:
\begin{equation}
\alpha \Delta(z)-\psi(\Delta(z))\le \epsilon
\end{equation}
from which it follows:
\begin{equation}\label{eq:Deltaz}
(\alpha-L_\psi) \Delta(z)\le \epsilon
\end{equation}
which proves the first inequality of (\ref{lachose}) with $\kappa_1=1/(\alpha-L_\psi)$, as $\alpha>L_\psi$.  
In order to prove the second inequality in (\ref{lachose}), we first rewrite $0 \le \ell(z) + \alpha\Delta(z) \le \epsilon$ as follows:
\begin{equation}
-\alpha\Delta(z)\le \ell(z)\le \epsilon \label{onf56ell}
\end{equation}
Combining \eqref{eq:Deltaz} together with (\ref{onf56ell}) implies that:
\begin{equation}
-[\dfrac{\alpha}{\alpha-L_\psi}]\epsilon\le \ell(z)\le \epsilon
\end{equation}
which proves that $\ell(z) \in \in [-\kappa_2\epsilon,\epsilon]$ with $\kappa_2:=\alpha/(\alpha-L_\psi)$.
$\hfill\Box$\\ \ \\ 
\begin{corollary}
Given a compact set $\mathbb Z:=\mathbb X\times \mathbb U$, if $\alpha$ is sufficiently high to meet the condition of Lemma \ref{lemmmmm}, there exists a continuous function $\varphi_s(\cdot)$, vanishing at $0$ such that for sufficiently small $\epsilon>0$, the following implication holds for all $z\in \mathbb Z$:
\begin{equation}
\Bigl\{\ell(z)+\alpha\Delta(z)\le \epsilon\Bigr\}\Rightarrow \Bigl\{d(z,\mathcal Z_s)\le \varphi_s(\epsilon)\Bigr\} \label{ygf76}
\end{equation}
\end{corollary}
{\sc Proof}. This is a straightforward consequence of Lemma \ref{lemmmmm} and Assumption \ref{ellbounded}. Indeed {\color{black} let us proceed by contradiction}, if (\ref{ygf76}) does not hold then {\color{black} it is possible to exhibit a} sequence of points $z^{(j)}\in \mathbb Z$ such that $\lim_{j\rightarrow\infty}\ell(z^{(j)})=0$ and  $\lim_{j\rightarrow\infty} \Delta(z^{(j)})=0$ while $\lim_{j\rightarrow\infty} d(z^{(j)},\mathcal Z_s)>r>0$ {\color{black} for some non-vanishing $r$}, which, by continuity and compactness argument leads to the existence of some $z^{(\infty)}$ that is steady optimal while lying outside $\mathcal Z_s$ which is obviously a contradiction by definition of $\mathcal Z_s$. $\hfill \Box$
\begin{lemma}[Properties of terminal pairs]\label{boundterminal}
Given any compact set $\mathbb X\subset \mathbb X_0$ of initial states and any $\alpha > L_{\psi}$, there exist 
two positive reals $\kappa_3, \kappa_4>0$ such that for any $x\in \mathbb X$, the optimal open-loop trajectory that solves the optimization problem (\ref{defdePx}) satisfies the following two terminal inequalities\footnote{Recall (\ref{notation*}) for the notation.}:
\begin{equation}
  \Delta_N^\star(x)\le \dfrac{\kappa_3(\alpha)}{\gamma}\qquad \mbox{\rm and}\qquad \vert \ell_N^\star(x)\vert\le \dfrac{\kappa_4}{\gamma} \label{ineqlem1}
\end{equation}
where $\kappa_3(\alpha)$ and $\kappa_4$ depend on $\mathbb X$ and $\mathbb U$. Moreover, the following asymptotic property holds true:
\begin{equation}
\lim_{\alpha\rightarrow \infty}\kappa_3(\alpha)=0 \label{kappa3vanishing}
\end{equation}
\end{lemma}
{\sc Proof}. Let us consider the following definitions:
\begin{align}
V^\star(x)&:= \sum_{k=0}^{N-1}\Bigl[\ell_k^\star(x)+\alpha\Delta_k^\star(x)\Bigr] \label{defdeSstar}\\
V^\#(x)&:= \sum_{k=0}^{N-1}\Bigl[\ell_k^\#(x)+\alpha\Delta_k^\#(x)\Bigr] \label{defdeSsharp}
\end{align}
corresponding to the sums of the stage costs over the trajectories starting from $x$ under the optimal control $\bm u^\star(x)$ and the control $\bm u^\#(x)$ invoked in reachability Assumption \ref{Ass_reach}. It comes by definition that:
\begin{align}
J(\bm u^\#,x)&=V(\bm u^\#, x) \quad \mbox{\rm \footnotesize (vanishing terminal costs)}\label{Jsharp}\\
J^\star(x)&:=V(\bm u^\star, x)+\gamma(\ell^\star_N(x)+\alpha \Delta^\star_N(x))
\end{align}
This gives by optimality of $J^\star(x)$:
\begin{equation}
\ell^\star_N(x)+\alpha \Delta^\star_N(x)\le \dfrac{V^\#(x)-V^\star(x)}{\gamma}\le \dfrac{K_0}{\gamma} \label{K0}
\end{equation}
for some positive constant $K_0 \le 2\max_{(x,u)\in \mathbb{X} \times \mathbb{U}^{N+1}} V(\bm u,x)$ which exists by virtue of the continuity of the involved maps (Assumption \ref{ellbounded}). Using inequality (\ref{K0}) together with Lemma \ref{lemmmmm} obviously gives the results for with $\kappa_3:=K_0\kappa_1$ and $\kappa_4=K_0\max\{1,\kappa_2\}$.  As for the asymptotic property (\ref{kappa3vanishing}), it comes directly from the fact that $\kappa_3=K_0\kappa_1$ and the result of Lemma \ref{lemmmmm} according to which $\kappa_1=1/(\alpha-L_\psi)$. 
$\hfill\Box$\\ \ \\ 
The following straightforward corollary is used in the proof of the main result:
\begin{corollary}\label{corcor}
	Under the Assumptions and notation of Lemma \ref{boundterminal}, the following inequality holds for all $x\in \mathbb X$:	
	\begin{equation}
	V^\star(x)\le V(\bm u^\# , x)+\kappa_4
	\end{equation}
\end{corollary}
{\sc Proof}. This is a direct consequence of the inequality:
\begin{equation}
V^\star(x)+\gamma(\ell_N^\star(x)+\alpha\Delta^\star_N(x))\le V(\bm u^\#, x)
\end{equation}
which, by virtue of (\ref{ineqlem1}), obviously implies that:
\begin{equation}
V^\star(x)\le V(\bm u^\#, x)-\gamma\ell^\star_N(x)\le V(\bm u^\#, x)+\kappa_4
\end{equation}
which proves the corollary. $\hfill \Box$ \\ \ \\ 
The following is another consequence of Lemma \ref{boundterminal} that is crucial in the proof of the main result:
\begin{corollary}[\small Recursive satisfaction of terminal properties] \label{correcursive}
	For any state on the closed-loop trajectory starting at $\mathbb X_0$, the inequalities (\ref{ineqlem1}) hold provided that $\alpha$ and $\frac{\gamma}{\alpha}$ are taken sufficiently high.
\end{corollary}
{\sc Proof}. Indeed since $\alpha$ is sufficiently high and the initial state lies in $\mathbb X_0$, the inequalities (\ref{ineqlem1}) hold for $x_0$. On the other hand, if $\gamma$ is taken sufficiently high these inequalities imply that $\Psi_N (\bm u^\star(x_0),x_0)=\eta$ with sufficiently small $\vert \eta\vert \le (\kappa_4+\alpha\kappa_3)/\gamma$. This makes the implication (\ref{condcindAss}) valid, then it can be deduced that there is $\bm u^\#:=\bm u^\dag$ that steers the next state  $x_{1}$ on the closed loop trajectory  to $\mathcal Z_s$, that is $x_1 \in \mathbb{X}_0$. 
This means that the arguments used in the proof of Lemma \ref{boundterminal} can be reused to show that (\ref{ineqlem1}) hold for the next state. By induction, the argument can now be reiterated to prove that these inequalities hold for all the states on the closed-loop trajectory. $\hfill \Box$

By now, we have all results we need to prove the main result of this paper.
\begin{prop} 
	[Main result] \label{Main}
For any desired $\epsilon>0$, provided that $\alpha$ is high enough, there exists sufficiently high $\gamma>0$ such that any resulting closed-loop trajectory starting at $x_0\in \mathbb X_0$ asymptotically becomes $\epsilon$-quasi steady optimal in the sense of Definition \ref{defregimquasi}. Moreover, for dynamics that are obtained by time sampling, the size of the terminal region might be further reduced  by reducing the sampling period $\tau$. 
\end{prop}
{\sc Proof}. Consider the following definition:
\begin{equation}
\bar V := \sup_{x_0\in \mathbb X_0} V(\bm u^\#(x_0), x_0) \label{defdeJbar}
\end{equation}
and the resulting set:
\begin{equation}
\mathbb X:=\Bigl\{x\in \mathbb{R}^n\ \vert\  V^\star(x)\le \kappa_4+\max\{\bar V,K_s(1)\}\Bigr\} \label{defdeBBX}
\end{equation}
where $K_s(\cdot)$ is the map invoked in Assumption \ref{localAss}. The subset $\mathbb X$ is bounded below by virtue of (\ref{lachose}). Now since by Corollary \ref{corcor}, $V^\star(x_0)\le \kappa_4+V^\#(x_0)\le \kappa_4+\bar V$, it comes out by induction that if it can be proved that when the $(x_k,u^\star_0(x_k))$ lies inside $\mathbb Z:=\mathbb X\times \mathbb U$ so is the next pair on the closed-loop trajectory, then the closed-loop trajectory remains inside $\mathbb Z$. Therefore, $\mathbb Z$-related Lipschitz constants can be invoked recursively. \\ \ \\ 
Using a standard receding horizon argument, {\color{black} the warm start sequence defined by  (\ref{ubfdag}) can be used as a candidate sequence for the optimization problem defined at $x_{k+1}$ and since the optimal solution has a lower cost than any admissible candidate solution}, it follows that at any state $x_k$ on the closed-loop trajectory
\begin{align}
V^\star(x_{k+1}) \le  &V^\star(x_k)-(\ell^\star_0(x_k)+\alpha\Delta^\star_0(x_k))+\Psi_N^*(x_k)+ \nonumber \\ 
&+\gamma\Bigl[\Psi(\bm u^{*+},x_{k+1})-\Psi^\star_N(x_k)\Bigr]\label{ineqqq}	
\end{align}
where
\begin{align*}
\Psi(\bm u^{*+},x_{k+1})&:=[\ell+\alpha\Delta](f(x^\star_N,u^\star_N),u^\star_N) \nonumber \\
&\le \Psi^\star_N(x_k)+(L_\ell+\alpha L_\Delta)[\tau\Delta^\star_N(x_k)]
\end{align*}
where $L_\ell$ and $L_\Delta$ are the Lipschitz constants of $\ell$ and $\Delta$ over $\mathbb X\times \mathbb U$. Using the first inequality (\ref{ineqlem1}) of Lemma \ref{boundterminal}, the last inequality can be rearranged to give:
\begin{equation}
\Psi(\bm u^{*+},x_{k+1})\le \Psi^\star_N(x_k)+(L_\ell+\alpha L_\Delta)\dfrac{\kappa_3(\alpha)\tau}{\gamma}
\end{equation}
Now using this last inequality in (\ref{ineqqq}) gives:
\begin{align}
V^\star(x_{k+1}) \le  &V^\star(x_k)-(\ell^\star_0(x_k)+\alpha\Delta^\star_0(x_k))+\nonumber \\ 
&+\Psi_N^*(x_k)+\kappa_5(\alpha)\tau\label{ineqqq2}	
\end{align}
where $\kappa_5(\alpha):=(L_\ell+\alpha L_\Delta)\kappa_3(\alpha)$. 
Moreover, we have by virtue of Lemma \ref{boundterminal}:
\begin{equation}
\Psi^\star_N(x_k)\le \dfrac{\kappa_3(\alpha)+\alpha \kappa_4}{\gamma}=:\dfrac{\kappa_6(\alpha)}{\gamma}\label{cellsi}
\end{equation}
Therefore, inequality (\ref{ineqqq2}) becomes:
\begin{equation}
V^\star(x_{k+1})\le V^\star(x_k)-(\ell^\star_0(x_k)+\alpha\Delta^\star_0(x_k))+\varphi(\alpha,\gamma,\tau) \label{uhg83}
\end{equation}
where 
\begin{equation}
\varphi(\alpha,\gamma,\tau):=\dfrac{\kappa_6(\alpha)}{\gamma}+\kappa_5(\alpha)\tau
\end{equation}
Consider the following set:
\begin{equation}
\mathcal A:= \Bigl\{x\ \vert\ \ell^\star_0(x)+\alpha\Delta^\star_0(x)\le 2\varphi(\alpha,\gamma,\tau)\Bigr\} \label{defdemathcalA}
\end{equation}
inequality (\ref{uhg83}) clearly shows that as long as the state $x_k$ on the closed-loop trajectory remains outside $\mathcal A$, $V^\star$ decreases at the next step keeping the closed loop trajectory inside $\mathbb X$. {\color{black} The amount of decrease is not vanishing since its amplitude is greater than $\varphi(\alpha,\gamma,\tau)$}. Now since $V$ is \rev{bounded} below over the compact set $\mathbb X$ {\color{black} for sufficiently high $\alpha$ and $\gamma/\alpha$}, this cannot occur indefinitely. Therefore, there exists a finite $\bar k$ such that:
\begin{equation}
\ell^\star_0(x_{\bar k})+\alpha\Delta^\star_0(x_{\bar k})\le 2\varphi(\alpha,\gamma,\tau)\quad \mbox{\rm and}\  x_{\bar k-1}\in \mathbb X
\end{equation} 
and using Assumption \ref{localAss}, this simply means that for sufficiently high $\alpha$ and $\gamma$, $x_{\bar k}$ satisfies $V^\star(x_{\bar k})\le \kappa_4+V^\#(x_{\bar k})\le \kappa_4+K_s(\varphi_s (2 \varphi(\alpha, \gamma,\tau)))\le \kappa_4+K_s(1)$. This means that $x_{\bar k}\in \mathbb X$ [see (\ref{defdeBBX})]. This clearly shows that the closed-loop state trajectory lies constantly inside $\mathbb X$ and the use of the Lipschitz constants is relevant and hence the resulting inequality (\ref{cellsi}) always holds.\\ \ \\ 
Now let us examine what happens for $k\ge \bar k$:
\begin{itemize}
\item Either $x_{k+1}$ remains in $\mathcal A$ in which case we have by definition of $\mathcal A$ and Lemma \ref{lemmmmm} that $\Delta(z_{k+1})\le 2\kappa_1\varphi(\alpha,\gamma,\tau)$ and $\vert \ell(z_{k+1})\vert \le 2 \max\{1,\kappa_2\}\varphi(\alpha,\gamma,\tau)$
\item Or $x_{\bar k+1}$ goes outside $\mathcal A$ but the resulting increase in $V^\star$ is limited by the fact that $\Delta^\star_0(x_{\bar k})\le 2\kappa_1\varphi(\alpha,\gamma,\tau)$ thanks to (\ref{lachose}) of Lemma \ref{lemmmmm}, before $V^\star$ decreases again (since the state is outside $\mathcal A$). 
\end{itemize} 
therefore, for all $k\ge \bar k$, the following inequalities hold:
\begin{align}
\vert \ell(x_k,u^\star_0(x_k))\vert&\le  2 \Bigl[\max\{1,\kappa_2\}+\kappa_1L_\ell\Bigr]\varphi(\alpha,\gamma,\tau)\\
\Delta(x_k,u^\star(x_k)) &\le 2\kappa_1(1+L_\Delta)\varphi(\alpha,\gamma,\tau)
\end{align}
which  proves the result since the above inequality means that the closed-loop trajectory is asymptotically $\epsilon$-quasi steady optimal in the sense of Definition \ref{defregimquasi} with $\epsilon$ vanishing as $\alpha$ and $\gamma/\alpha$ tend towards infinity. As for the role of decreasing $\tau$ inequality (\ref{uhg83}) clearly shows that smaller values of $\tau$ induce smaller terminal set $\mathcal A$ given by (\ref{defdemathcalA}). $\hfill \Box$

\subsection{Discussion on state constraints}\label{secsoft}
{\color{black} In this section, the soft constraint issue is discussed in more details having in mind the facts that have been just proved regarding the behavior of the closed-loop system. This discussion is not meant to be a rigorous proof. Rather, it gives hints regarding possible theoretical argumentation that we avoided for the sake of brevity and in order to convey the main ideas. \\ \ \\ We assume that an exact penalty of the form $\rho \max\{0,g(x)\}$ is included in the stage cost to soften the constraints $g(x)\le 0$. We assume that the optimal steady pair lies in the interior of the admissible set. Moreover, we assume that the definition of the set $\mathbb X_0$ is now restricted to those initial states for which an admissible reachability trajectory exists. \\ \ \\ Based on these assumptions, it can be easily proved (as it has been done in Lemma \ref{boundterminal}) that for any such initial state, the total cost is bounded by an upper bound that does not depend on $\rho$ since the admissible trajectory has no constraint violation cost. This means that the amount by which the initial optimal trajectory violates the constraint is bounded by a term that behaves as $O(1/\rho)$. On the other hand, if $\alpha$ and $\gamma$ are high enough, the open-loop optimal trajectory ends inside the admissible region meaning that the recursive feasibility argument holds. 
Then, the decreasing properties of the cost function established in the proof of Proposition \ref{Main} are valid until the closed-loop trajectory reaches the small $\epsilon$-neighborhood of the steady pair which is inside the admissible region. This simply means that over the closed-loop trajectory, the violation of the constraints will be constantly lower than an amount that tends to $0$ as $\rho$ increases.}
\section{Illustative example}\label{secExamples}
For the sake of illustration and to make an easy comparison with literature, let us consider the commonly used example of the nonlinear continuous flow stirred-tank reactor with parallel reactions \cite{Bailey:1971}.
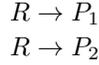
\begin{align*}
&R\rightarrow P_1\\
&R\rightarrow P_2
\end{align*}
that can be described by the following dimensionless energy and material balances:
\begin{subequations}
	\begin{align}
\dot x_1&=1-10^4x_1^2e^{-1/x_3}-400x_1e^{-0.55/x_3}-x_1 \label{reac1}\\
\dot x_2&=10^4x_1^2e^{-1/x_3}-x_2 \label{reac2}\\
\dot x_3&=u-x_3 \label{reac3}
\end{align}
\end{subequations}
where $x_1$ and $x_2$ stand for the concentrations of $R$ and $P_1$ respectively while $x_3$ represents the temperature of the mixture in the reactor. $P_2$ represents the waste product. The control variable is given by the heat flow $u\in [0.049, 0.449]$. The natural stage cost would be given by $\ell(x,u)=-x_2$ since the objective is to maximize the amount of product $P_1$.\\ 

Different aspects of using EMPC to address this problem have been considered in \cite{Muller:2014} where it has been recalled that without any particular care, there is a non stationary optimal solution to the purely economic formulation that would include only the stage cost $\ell=-x_2$ while the use of average constraint enable to reduce the level of oscillations in the closed-loop behavior. It has been also shown that the system possesses an optimal steady pair denoted hereafter by $x_s=(0.0832, 0.0846, 0.149)$ and $u_s=0.149$. \\ \ \\ {\color{black} In the present section, the behavior of the closed-loop under the EMPC associated to the proposed formulation is analyzed for different choices of the design parameters in order to assess the underlying theoretical development. In all the forthcoming simulations, the economic MPC design uses a prediction horizon length of $N=20$ for a prediction \rev{time-horizon of $2$} (this corresponds to a dimensionless sampling period of 0.1 used inside the predictor which is to be distinguished from the closed-loop control updating which is taken in $\{0.1, 0.02\}$) depending on the control settings. The optimization  was done using the CasADi-Python software running the IPOpt solver \cite{Andersson:2018} with a single shooting implementation. The maximum number of iterations has been fixed to $1000$. mEight control settings are investigated in this section in order to illustrate the impact of the parameters choice. They are split into two subsets that are investigated in Figures \ref{fig1} and \ref{fig2} respectively. \\ 

\begin{figure}
\begin{center}
\input{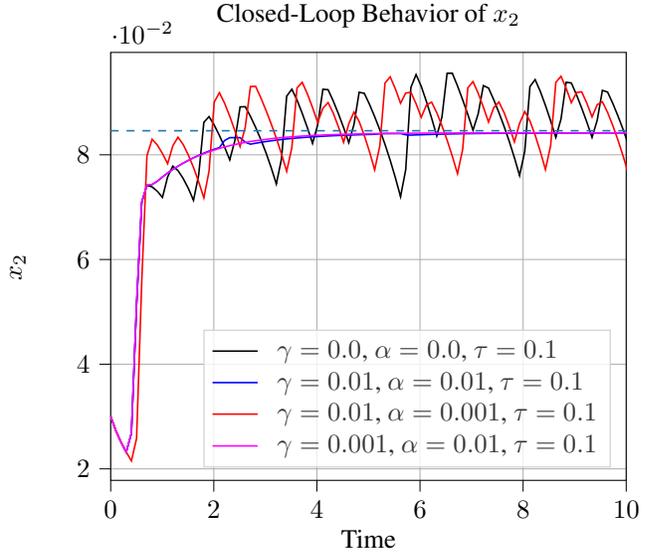}
\end{center}	
\caption{\color{black} Impact of $\alpha$ in stabilising the trajectories near the steady optimal value.} \label{fig1}
\end{figure}

\begin{figure}
\begin{center}
\input{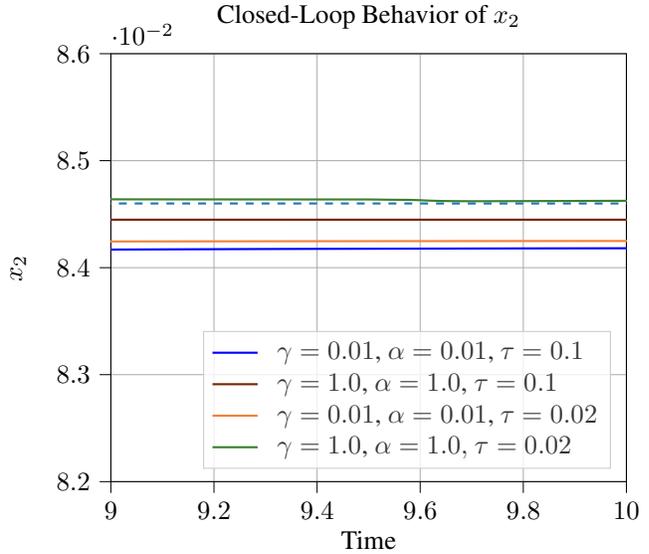}
\end{center}	
\caption{\color{black} Illustration of the increasing precision of the terminal regime with increasing values of $(\gamma,\alpha)$ and decreasing of values of $\tau$} \label{fig2}
\end{figure}

Figure \ref{fig1} shows the non-stationary behavior of the pure economic MPC formulation (black line) which corresponds to $\gamma=\alpha=0$. It also shows that if $\alpha$ is not high enough ($\alpha=0.001$), the stationarity is not achieved (red line) while for sufficiently high value of $\alpha=0.01$, a stationary final regime is obtained even for small values of $\gamma=0.001$.   
\\ \ \\ 
Figure \ref{fig2} investigates different settings for which only the tail of the closed-loop trajectory is shown in order to focus on the terminal precision being achieved. The first two plots (blue and brown) correspond to the same sampling period $(\tau=0.1)$ so that it can be clearly shown that higher values of $\gamma$ and $\alpha$ (here $\gamma=\alpha=1$) induce higher precision than the lower values setting (here $\gamma=\alpha=0.01$). \\ 
The other remaining curves (green and orange) represent the results for the previous two settings for which a smaller sampling period $\tau=0.02$ is considered in order to illustrate the impact predicted by Proposition \ref{Main}.\\ \ \\ 
Finally, Figure \ref{fig3} investigates the impact of taking $\gamma=0$ in the formulation. The result confirms what is partially suggested by the results of Figure \ref{fig2}, namely that for this specific example, it is mainly $\alpha$ that plays the major role despite the fact that $\gamma$ is needed in the proof of closed-loop stability. Nonetheless, note that, as in all provably stable MPC frameworks, the underlying conditions are only sufficient but not necessary. It can be noticed, however, that the convergence is very slightly accelerated by the use of $\gamma=1$ but the effect is probably too small to be solidly assessed. 
}
\begin{figure}
\begin{center}
\input{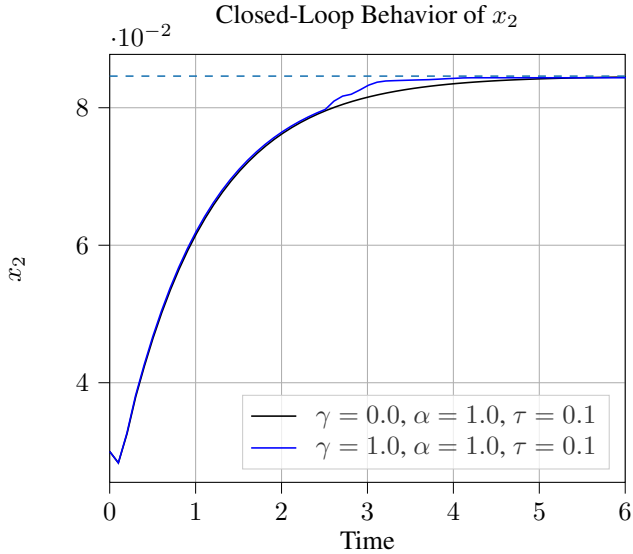}
\end{center}	
\caption{\color{black} Impact of $\gamma$.} \label{fig3}
\end{figure}

\section{Conclusion} \label{secconc}
In this paper, a new formulation of economic MPC is proposed for discrete-time dynamics. 
The formulation needs no terminal constraints on the state and is based on the penalization of the state increments between two successive states.
Convergence to a quasi optimal steady regime has been derived using rather mild technical conditions, and the distance to the optimal equilibrium can be made as small as desired by changing the design parameters for fixed prediction horizon, which only needs to be compatible with a reachability assumption.

\bibliographystyle{plain}
\bibliography{param_alamir.bib}

\begin{thebibliography}{10}

\bibitem{Muller:2014}
M\"{u}ller~M. A., David Angeli, Frank Allg\"{o}wer, Rishi Amrit, and James~B. Rawlings.
\newblock Convergence in economic model predictive control with average constraints.
\newblock {\em Automatica}, 50(12):3100 -- 3111, 2014.

\bibitem{Amrit:11}
R.~Amrit, J.B. Rawlings, and D.~Angeli.
\newblock Economic optimization using model predictive control with a terminal cost.
\newblock {\em Annual Reviews in Control}, 35(2):178 -- 186, 2011.

\bibitem{amrit2013}
Rishi Amrit, James~B Rawlings, and Lorenz~T Biegler.
\newblock Optimizing process economics online using model predictive control.
\newblock {\em Comp. \& Chem. Eng.}, 58:334--343, 2013.

\bibitem{Angeli:12a}
D.~Angeli, R.~Amrit, and J.B. Rawlings.
\newblock On average performance and stability of economic model predictive control.
\newblock {\em IEEE Trans. Automat. Contr.}, 57(7):1615--1626, 2012.

\bibitem{Diehl:11a}
M.~Diehl, R.~Amrit, and J.~B. Rawlings.
\newblock A {L}yapunov function for economic optimizing model predictive control.
\newblock {\em IEEE Trans. Automat. Contr.}, 56:703--707, 2011.

\bibitem{Andersson:2018}
Andersson J.~A. E., J.~Gillis, G.~Horn, J.~B Rawlings, and M.~Diehl.
\newblock {CasADi} -- {A} software framework for nonlinear optimization and optimal control.
\newblock {\em Mathematical Programming Computation}, 2018.

\bibitem{Bailey:1971}
Bailey~J. E., F.~J.~M. Horn, and R.~C. Lin.
\newblock Cyclic operation of reaction systems: effect of heat and mass transfer resistance.
\newblock {\em {AIC}h{E} Journal}, 17(4):818--825, 1971.

\bibitem{faulwasser:pannocchia:2019}
Timm Faulwasser and Gabriele Pannocchia.
\newblock Toward a unifying framework blending real-time optimization and economic model predictive control.
\newblock {\em Industrial \& Engineering Chemistry Research}, 58(30):13583--13598, 07 2019.

\bibitem{Griffith17a}
D~W Griffith, Vr~M Zavala, and L~T Biegler.
\newblock Robustly stable economic nmpc for non-dissipative stage costs.
\newblock {\em Journal of Process Control}, 57:116--126, 2017.

\bibitem{Gruene13a}
L.~Gr{\"u}ne.
\newblock Economic receding horizon control without terminal constraints.
\newblock {\em Automatica}, 49(3):725--734, 2013.

\bibitem{maree2016}
Johannes~Philippus Maree and Lars Imsland.
\newblock Combined economic and regulatory predictive control.
\newblock {\em Automatica}, 69:342--347, 2016.

\bibitem{qin:badgwell:2003}
S.~J. Qin and T.~A. Badgwell.
\newblock A survey of industrial model predictive control technology.
\newblock {\em Control Engineering Practice}, 11:733--764, 2003.

\bibitem{Rawlings09b}
J.~B. Rawlings and R.~Amrit.
\newblock Optimizing process economic performance using model predictive control.
\newblock In L.~Magni, D.~Raimondo, and F.~Allg{\"o}wer, editors, {\em Nonlinear Model Predictive Control - Towards New Challenging Applications}, volume 384 of {\em Lecture Notes in Control and Information Sciences}, pages 119--138. Springer Berlin, 2009.

\bibitem{rawlings:mayne:diehl:2017}
James~B. Rawlings, David~Q. Mayne, and Moritz~M. Diehl.
\newblock {\em Model Predictive Control: Theory, Computation, and Design}.
\newblock Nob Hill Publishing, Madison, WI, second edition, 2017.

\bibitem{kit:zanon18a}
M.~Zanon and T.~Faulwasser.
\newblock Economic {MPC} without terminal constraints: Gradient-correcting end penalties enforce stability.
\newblock {\em Journal of Process Control}, 63:1--14, 3 2018.

\bibitem{zavala2015}
Victor~M Zavala.
\newblock A multiobjective optimization perspective on the stability of economic mpc.
\newblock {\em IFAC-PapersOnLine}, 48(8):974--980, 2015.

\end{thebibliography}

\end{document}